\documentclass[10pt, twocolumn, a4paper]{scrartcl}

\pdfoutput=1
\usepackage[cmex10]{amsmath}
\usepackage{amssymb}
\usepackage{a4wide}
\usepackage{balance}

\usepackage[caption=false,font=footnotesize]{subfig}

\usepackage{fixltx2e}
\usepackage{graphicx}

\usepackage{color}
\usepackage{bm}
\usepackage{nameref}
\usepackage[latin1]{inputenc}
\usepackage{multirow}
\usepackage{booktabs}
\usepackage{tabularx}
\usepackage{tikz}
\usepackage{pgfplots}
\usepackage{xparse}
\usepackage{ifthen}
\usepackage{url}

\pgfplotsset{compat=1.5}

\newcommand{\vm}[1]{\protect\ensuremath\boldsymbol{#1}} 
\newcommand{\rfs}[1]{\mathrm{#1}} 

\newcommand{\prio}{{k-1}}
\newcommand{\pred}{{k|k-1}}
\newcommand{\post}{{k}}
\newcommand{\kp}{{k}}

\newcommand{\birth}{b}

\newcommand*\xbar[1]{%
  \hbox{%
    \vbox{%
      \hrule height 0.5pt 
      \kern0.2ex
      \hbox{%
        \kern-0.02em
        \ensuremath{#1}%
        \kern-0.02em
      }%
    }%
  }%
} 

\NewDocumentCommand{\var}{O{} m O{}}{
  \ifthenelse{\equal{#3}{\empty}}
	  {\ensuremath{#2_{#1}^{}} }
		{\ensuremath{#2_{#1}^{#3}} }
}
\NewDocumentCommand{\varNoAlign}{O{} m O{}}{
  \ifthenelse{\equal{#3}{\empty}}
	  {\ensuremath{#2_{#1}} }
		{\ensuremath{#2_{#1}^{#3}} }
}
\newcommand{\x}[1][\empty]{
  \var{\vm{x}}[#1]
}

\newcommand{\xprio}[1][\empty]{
  \var[\prio]{\vm{x}}[#1]
}
\newcommand{\xpred}[1][\empty]{
  \var[\pred]{\vm{x}}[#1]
}
\newcommand{\xpost}[1][\empty]{
  \var[\post]{\vm{x}}[#1]
}

\newcommand{\z}[1][\empty]{
  \var{\vm{z}}[#1]
}

\newcommand{\prob}{
	p
}

\newcommand{\Zmeas}{\var[\kp]{\rfs{Z}}}

\newcommand{\spatialClutterDensity}[1][\empty]{
  \var{\varkappa}[#1]
}

\newcommand{\pDetect}[1][\empty]{
  \var[D]{p}[#1]
}

\newcommand{\pSurvive}[1][\empty]{
  \var[S]{p}[#1]
}
\newcommand{\birthDensity}[1][\empty]{
  \var[\prio]{\birth}[#1]
}

\newcommand{\intensity}[1][\empty]{
  \var{v}[#1]
}

\newcommand{\meanClutter}{\lambda}

\newcommand{\SigmaPoint}[1][\empty]{
  \mathcal{X}
}

\newcommand{\BetP}[1][\empty]{
 \ifthenelse{\equal{#1}{\empty}}
	  {\var[m]{BetP}}
		{\var[#1]{BetP}}
}
\newcommand{\Bel}[1][\empty]{
 \ifthenelse{\equal{#1}{\empty}}
	  {\var[m]{Bel}}
		{\var[#1]{Bel}}
}
\newcommand{\Pl}[1][\empty]{
 \ifthenelse{\equal{#1}{\empty}}
	  {\var[m]{Pl}}
		{\var[#1]{Pl}}
}
\newcommand{\Uncertainty}[1][\empty]{
 \ifthenelse{\equal{#1}{\empty}}
	  {\var[m]{U}}
		{\var[#1]{U}}
}

\begin{document}
\title{Information Maps: A Practical Approach to Position Dependent Parameterization}
\author{Benjamin Wilking, Daniel Meissner, Stephan Reuter, \\and Klaus Dietmayer\\
Institute of Measurement, Control, and Microtechnology\\
Ulm University, Germany\\
\begin{small}\{benajmin.wilking, daniel.meissner, stephan.reuter, klaus.dietmayer\}@uni-ulm.de\end{small}}
\date{}

\maketitle

\begin{abstract}
In this contribution a practical approach to determine and store position dependent parameters is presented. These 
parameters can be obtained, among others, using experimental results or expert knowledge and are stored in 
'Information Maps'. Each Information Map can be interpreted as a kind of static grid map and the framework allows to link 
different maps hierarchically. The Information Maps can be local or global, with static and dynamic information in it. 
One application of Information Maps is the representation of position dependent characteristics of a sensor. Thus, 
for instance, it is feasible to store arbitrary attributes of a sensor's preprocessing in an Information Map and utilize them 
by simply taking the map value at the current position. This procedure is much more efficient than using the attributes of the 
sensor itself.
Some examples where and how Information Maps can be used are presented in this publication. The Information Map is meant to be a 
simple and practical approach to the problem of position dependent parameterization in all kind of algorithms when 
the analytical description is not possible or can not be implemented efficiently.
\end{abstract}

\section{Introduction}
\subsection{Motivation}
Today, there are many different tasks to proceed in topics like advanced driver assistance systems or highly 
autonomous driving. These tasks can be e.g. sensor preprocessing including classification problems or multi-object 
tracking algorithms. There is a huge amount of different algorithms to solve all the challenging problems coming up 
with advanced assistance or autonomous systems. Coming along with all these algorithms is one widespread purpose: the 
parameterization. Obviously, the spatial resolution of common automotive sensors, like cameras, radars, and laser 
range finders decrease with the radial distance. But, especially algorithms for object detection highly depend on the 
sensors resolution. To model these dependencies as parameters analytically is not always possible. One of these 
parameters is the detection probability of every sensor in a multi-object tracking algorithm. If this parameter is 
not modeled correctly, it might happen, that information about an object from different sensors is ambiguous, e.g. a 
new track is initialized based on the measurement of one sensor, but a second sensor has no information about the new 
object. Because of this ambiguity, the track might be deleted again. Another example for the importance of modeling 
sensors is surround tracking for autonomous driving functions. It is not always possible to have sensors all around 
the own vehicle to observe the complete surrounding area without any blind spots. So the task is to model these blind 
spots, if possible even sensor independent. 
Adapting the tracking parameters is also a simple way to include information from different sources, like dynamic 
grid maps \cite{Thrun2005} \cite{Weiss2008} or digital (street) maps, into the tracking. At this point it is 
important to mention, that it is not objective of this contribution to completely avoid expert knowledge or 
heuristics, but to find a clean and practical approach to include such knowledge into algorithms like multi-object 
tracking. 

\subsection{Problem addressed}
Referring to the tracking example, without going into detail, the prediction and update equations of the Probability Hypothesis Density Filter (PHD)\cite{Mahler2007} are given by:
\begin{align}
\intensity(\xpred) = \int & \pSurvive(\xprio) \prob(\xpred|\xprio) \nonumber
\\ &\intensity(\xprio) d\xprio + \birthDensity(\x),
\end{align}
\begin{align}
&\intensity(\xpost) = \left[ 1-\pDetect(\x) \right] \intensity(\xpred) + \nonumber\\
& \sum_{\z \in \Zmeas} \frac{ \pDetect(\x) \prob(\z|\x) \intensity(\xpred)} {\meanClutter \spatialClutterDensity(\z) + \int{\pDetect(\xi) \prob(\z|\xi) \intensity(\xi_{\pred}) d\xi}}.
\end{align}
The necessary parameters here are: the detection probability $p_D$, the PHD of the new objects $b$, the persistence 
probability $p_S$ and the clutter probability $\varkappa$. Even the measurement uncertainties or the field of view of 
a sensor is not always as well-known as they are meant to be. That is the reason why many of these parameters became 
tuning parameters and are often set to a constant value or to an estimated distribution. The clutter probability for 
example is often assumed to be uniformly distributed within the gating volume. This contribution proposes to 
determine these parameters in practical experiments using static digital maps, called 'Information Maps'. Using such 
maps allows to analyze different sensors including their \emph{preprocessing} to get a more complete description of 
the perception performance. Furthermore, it is possible to incorporate dynamic information like already existing 
tracks. The Information Map approach also gives the ability to incorporate contextual information. Relating to the 
advanced driver assistance or to autonomous driving the Information Map is a generic and sensor independent way to 
handle multi-sensor fusion. But the information maps are not limited to represent tracking parameters. Other examples 
where useful, position dependent knowledge can be stored in maps are the a priori class probability of an Bayesian
classifier (see Section \ref{sec:ContextInformation}) and the maximum search radius of a clustering algorithm in 2D, 
etcetera.
\subsection{Related work}
Modeling the detection probability using a map was already done in different publications. One of them is 
\cite{Reuter2011}, in which the detection probability was modeled using an occupancy grid map in a static scenario. Two 
laser range finders are used to generate measurements and the detection probability dynamically depends on the 
occlusions in the scene. The basic idea to use a map to determine detection probabilities is quite similar to this 
contribution, but modeling the environment in different maps, like global or local/sensory, or with linked maps even 
including context information is not done in \cite{Reuter2011}. Another approach to solve occlusions using the 
detection probability was presented from \cite{Lamard2012}. Therein a method was presented to create a detection 
probability map built by convolving the target position and a width function. The main task of \cite{Lamard2012} is 
the same as in \cite{Reuter2011}, but incorporating the actual object dimensions and its uncertainties. This 
contribution does not compare to the way of modeling detection probabilities presented in \cite{Lamard2012}. In fact, 
the presented approach here allows to combine the information obtained from \cite{Lamard2012} with other sources. 
Further related work can be found in \cite{Lamard2012}.

This contribution is organized as follows: First, the 'Information Map' approach is explained in detail. Afterwards, 
a method to create static maps is presented, with a focus on detection probabilities for multi-object trackers. An 
excursion to how context information can be incorporated using Information Maps is followed by a short conclusion. 

\section{Information Map}
\begin{figure}	
	\def\svgwidth{0.9\columnwidth}
	\centering
	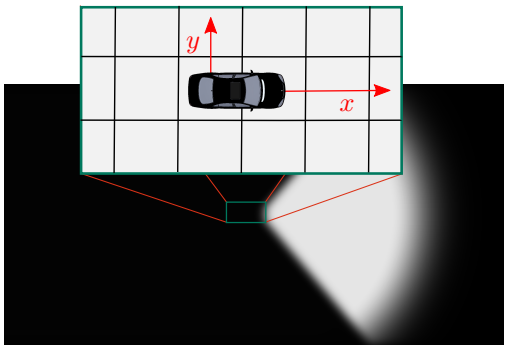
	\caption{Vehicle coordinate system. In the background: Example of a Detection Map with a resolution of $0.1m$.}
	\label{fig:CoordSystem}
\end{figure}

\begin{figure}	
	\def\svgwidth{0.9\columnwidth}
	\centering
	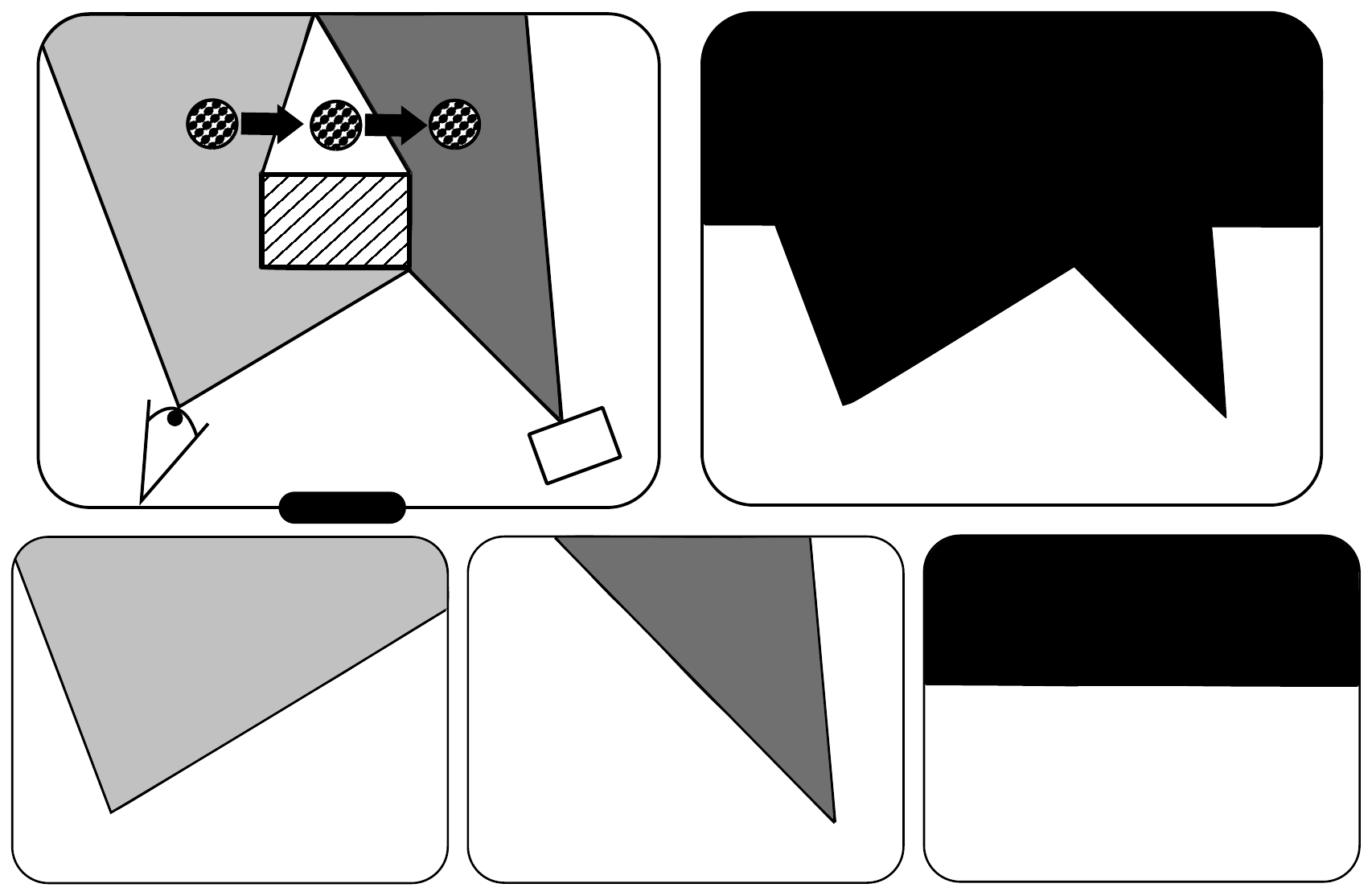
	\caption{Simplified example for the use of combining different maps: The upper left image shows the scene, where a 
tracked object crosses the occlusion. The lower left and center local maps show the sensor field of view and the 
lower right shows an global persistence map with the assumption, that object can not disappear in the upper part of 
the map, because there is only one door, where an object could enter or leave, between the two sensors. The combined 
global persistence map is in the upper right image. If the object enters the occluded area, the track does not get 
lost, because the persistence for this area is still high. A hierarchical request is not intended here, because the 
value of the occupied cells of the combined global map is not the same as in the local ones (coded by color).
}
	\label{fig:ExampleGlobal}
\end{figure}

\begin{figure}	
	\def\svgwidth{0.9\columnwidth}
	\centering
	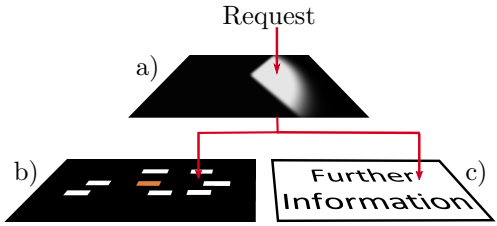
	\caption{Example of a hierarchical request to get a birth probability at the position of a certain measurement. 
First the birth probability is requested from the static local/sensory Birth Probability Map a). This map forwards 
the request to a global dynamic Object Map b), where all dynamic objects were inserted, and to any other Information 
Map appended c). In a last step the values from all maps are combined and the result is returned.}
	\label{fig:BirthProbExample}
\end{figure}

\balance
As already mentioned, this contribution proposes to use 'Information Maps' to store parameters. Therefore, the 
determined parameters are not continuous functions: they are discretized and stored as a kind of grid map. To do 
that, these maps are represented as large matrices. A easy way to work with matrices fast and effectively is e.g. the 
open source C++ library armadillo \cite{Sanderson2010}. So far, every map is a static map located at the origin of 
the tracking system, thus every map needs to know its own resolution and the origin of the local system in map 
coordinates (rows, columns) as shown for the tracking example in Fig. \ref{fig:CoordSystem}. 
The only limitation here is that the data is stored as a matrix. Thus it is also possible to save parameters in non 
Cartesian coordinates like the polar coordinate system. It is also necessary to differ between local and global maps. 
Referring to the tracking example, this means that every sensor can have its own local map for a certain tracking  
parameter. But sometimes it is also useful to have a global map incorporating all sensor specific information. In 
this context global means that the map is located at the origin of the tracking system like local/sensory maps, but 
it is valid for the complete system rather than limited to only one sensor. 
Short example:Two sensors observe an area with an occlusion. This occlusion is modeled using a Information Map. 
Tracking objects in the observed area suffers from the occlusion and the resulting track loss. If there is the 
knowledge, that the object crossing the occlusion can not disappear, the persistence probability in the area of 
occlusion can kept high. With that model using a combined global map from both sensors and the occlusion map the 
track will survive when crossing the occluded area (Fig. \ref{fig:ExampleGlobal}).
 In other cases, it is useful to have only one global and no local ones. Another property of Information Maps is the 
possibility of a hierarchical request. This means, that multiple maps can be linked together in an hierarchical 
manner. If there is an information request to a certain Information Map, the map does not only return it's own 
information, but the combined information of itself and all it's appended maps. An example for an hierarchical 
request is shown in Fig. \ref{fig:BirthProbExample}. Further, every map can be regarded as kind of interface between 
the matrix the information is stored in and the application which requests the information. Thus, an information map 
can also be an interface to other toolboxes and information sources like a dynamic grid map. Therefor, it is not 
necessary to extract the information from the source and save it into a new matrix. The Information Map only gets a 
position in local coordinates and delivers the result combining all appended Information Maps at this position.

Knowledge about the static behavior around the system is only half of the medal. In many cases it is also necessary 
to know something about the dynamic environment. A good example for that is the birth probability: it is not only 
helpful to know the perception range of a sensor, it is also useful to know all dynamic objects already existing as 
track hypotheses. Therefore not only static information is stored as a map. Consequently it follows, that certain 
parameters are not only determined using one single map, but rather using a set of different static and dynamic maps. 
This is realized using the already mentioned hierarchical requests (Fig. \ref{fig:BirthProbExample}).

\section{Creating static sensory maps}

\begin{figure}[ht!]
\centering
\def\svgwidth{0.8\columnwidth}
\subfloat[Three laser range finders mounted at the front of a vehicle.]{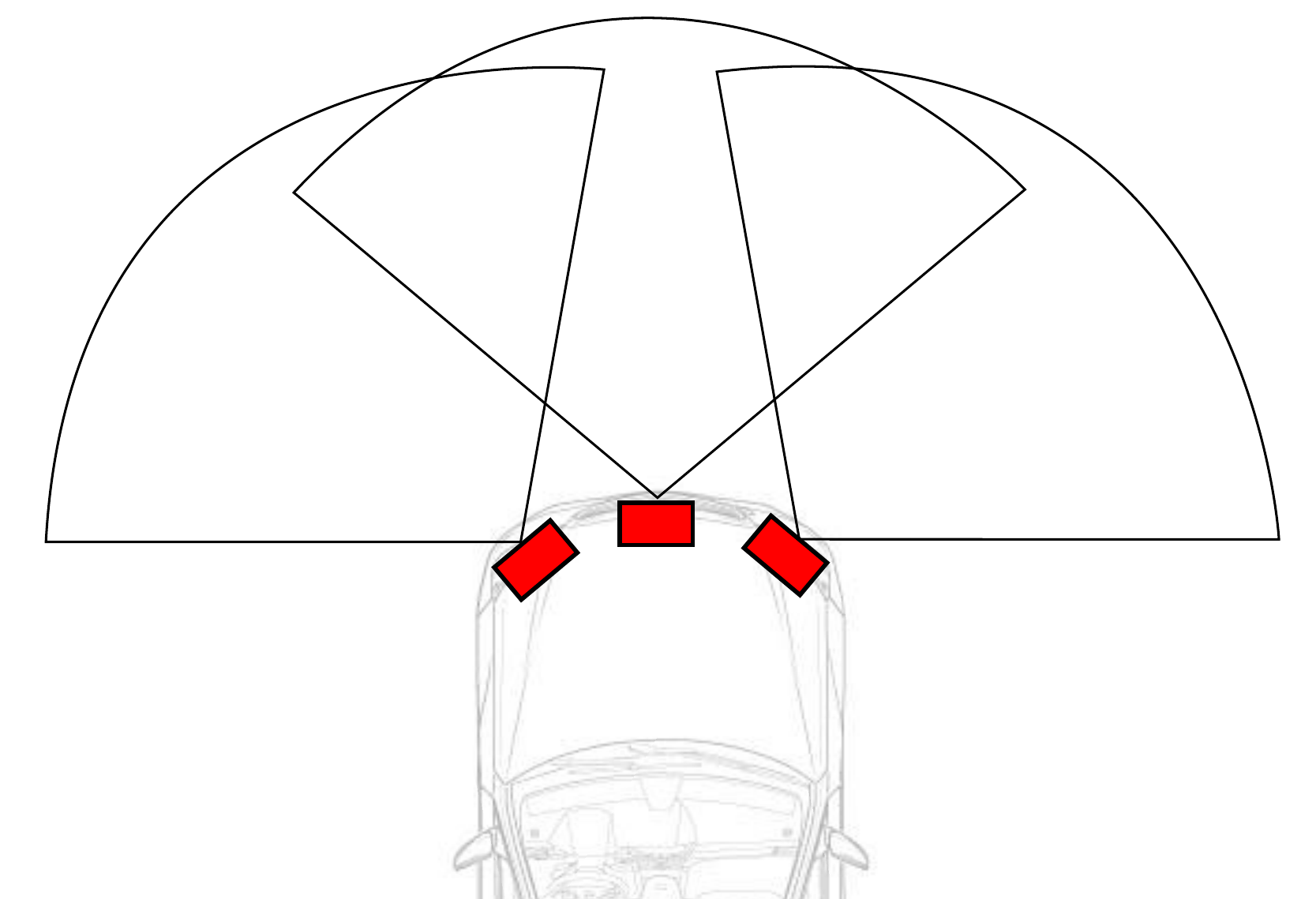 \label{fig:LSMount}}\\
\subfloat[Detections of the combined laser\newline range finders.]{{\input{pics/3LaserPoints}\hspace*{12mm}} \label{fig:TruePositiveDetections}}\\
\subfloat[Resulting Detection Map.]{\includegraphics[height=0.7\columnwidth,angle=90]{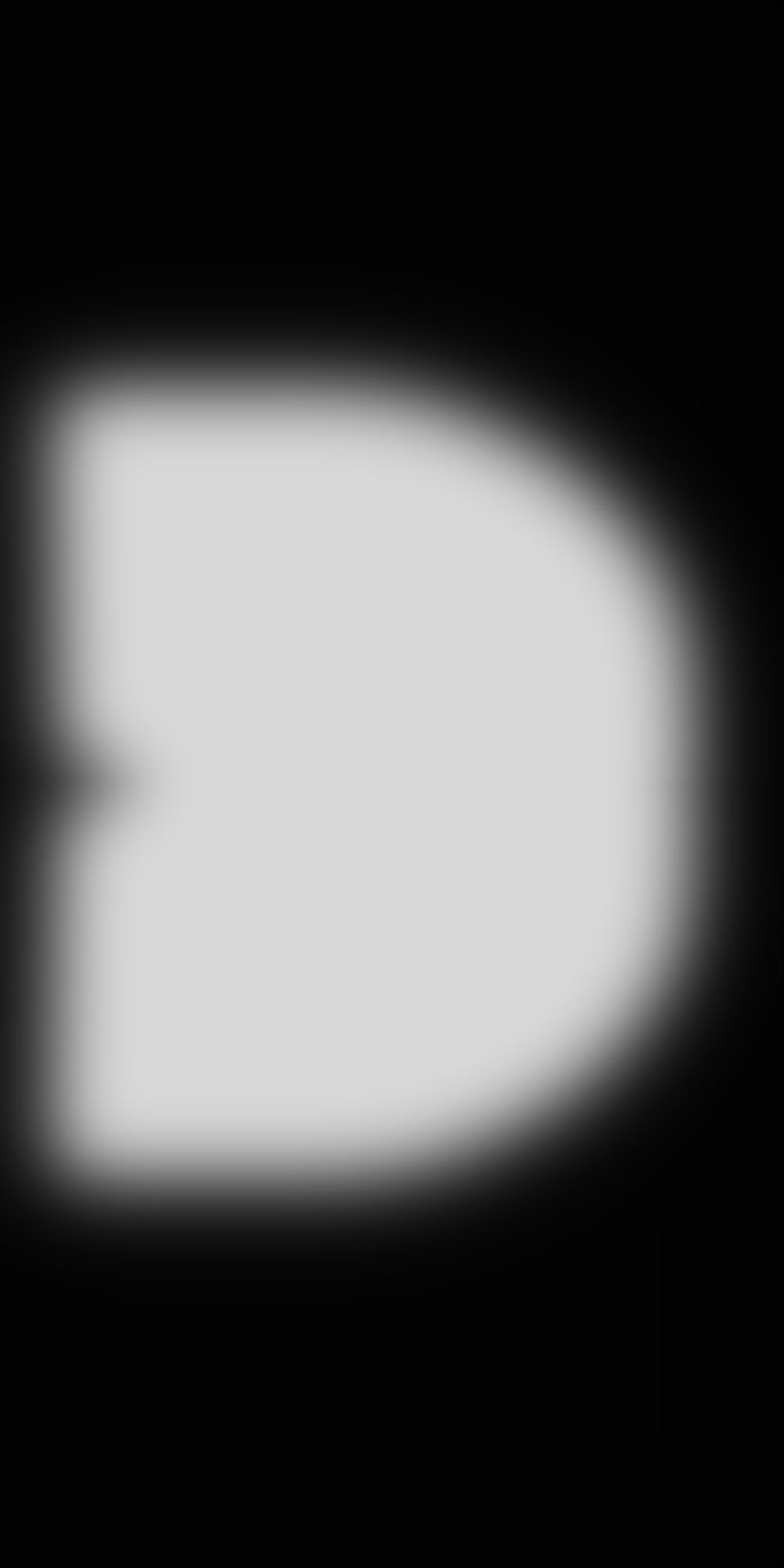} \label{fig:DetectionMap3Lasers}}\\
\caption{Scenario with three laser range finders mounted at the front of a vehicle. a) The mounting positions of the 
sensors. b) The measurements of the preprocessing, combining all three sensors in one virtual sensor. 
c) The resulting Detection Map.}
\label{fig:Scenario3Lasers}
\end{figure}

One big advantage of using Information Maps containing the attributes of a sensor is the possibility to determine the 
perception performance of the preprocessing of this sensor. These attributes are needed very often, 
but in most cases, these attributes are not known or wrongly assumed. This problem can be explained by a simple 
example using laser range finders: The laser range finder has a huge perception range and a relatively wide opening 
angle. But normally not the laser detections itself are used as measurements. In case of detecting 
vehicles, a preprocessing algorithm as described in \cite{Munz2009a}, where a box is fitted into the data of the 
laser range finder, can be used. But fitting boxes into data depends on the angular resolution of the laser scanner. 
Objects in a higher distance have fewer points than objects close to the range finder. Therefore the preprocessing 
has attributes completely different to those of the sensor itself.
A further problem occurs when combining sensors in a preprocessing step. The opening angle and range can't be 
determined anymore, when using multiple sensors as one 'virtual' sensor. Fig. \ref{fig:LSMount} shows an example 
where three laser range finders are mounted at three different positions at the front of a car. The data of all three 
range finders are transformed into the vehicle coordinate system and are used for the box fitting algorithm. Now there 
is one single 'virtual' sensor and therefore the need to know the attributes of this new sensor, 
respectively it's preprocessing, arises.\\
A good way to create a new map for a certain parameter is to analyze the preprocessing over time, which is 
illustrated by the following example: creating a static map for the detection probability of the 'virtual' sensor 
from above can be done as follows: after recording multiple or long sequences in different environments, e.g. 
motorway, country road, city, and so on, all detections, if possible true positives only, are plotted in one image 
(Fig. \ref{fig:TruePositiveDetections}). With that plot the perception range can be estimated. If ground truth 
data is available the percentage of detected vehicles, the detection probability, can be calculated for every cell 
of the map. In most cases it is not possible to calculate the complete map because the training data normally is not 
sufficient or there is no ground truth data available. Thus expert knowledge is needed to create the final map from 
the plotted data. A very practical approach for that is to use a simple image editing tool and to save the map in a 
standard image format. The benefit is that this map can easily be edited and afterwards it is very simple to load such
an image as an Information Map using e.g. OpenCV \cite{Bradski2000}. The result after incorporating the expert knowledge
can be seen in Fig. \ref{fig:DetectionMap3Lasers}. 
The very low detection probability right in front of the vehicle is caused by bad results of the box fit when the 
detected object is to close to the own vehicle. If ground truth data for the true positives is available it is also 
possible to determine the measurement uncertainty at every position and an 'Uncertainty Map' for the sensor as well 
as a 'Clutter Map' can be created. Such maps do not have to follow any probability distribution. Thus, it is possible 
to depict nearly every distribution for most sensors. After creating a static Detection Map it can be combined e.g. 
with the dynamic detection probability from \cite{Lamard2012}.
When using a sensor with measurements which can't be transformed into the vehicle coordinates, e.g. a camera, special 
a priori knowledge is needed. In case of a camera a video classifier like the Viola-Jones cascade \cite{Viola2001} 
can be used to detect vehicles. The transformation of the measurements into vehicle coordinates can be done assuming 
that the size of the objects is known or with the knowledge that the world around the vehicle conforms to the flat 
world assumption. This is necessary because of the loss of information in the third dimension, when projecting the 3D 
world to a 2D image. For the camera it is now possible to create a $p_D$ map in the same way as described for the 
laser range finders.
It is often assumed that the field of view is equal to the opening angle of the camera and the range is limited by the 
minimum size of a detection in pixels. But regarding that measurements created by a detector do not depend on the 
camera itself, but rather on the preprocessing, it is a better way to use the attributes of the preprocessing instead 
of the attributes of the camera itself.

\section{Context Information}
\label{sec:ContextInformation}
\begin{figure}[ht!]
	\centering
	\includegraphics[width=0.6\columnwidth]{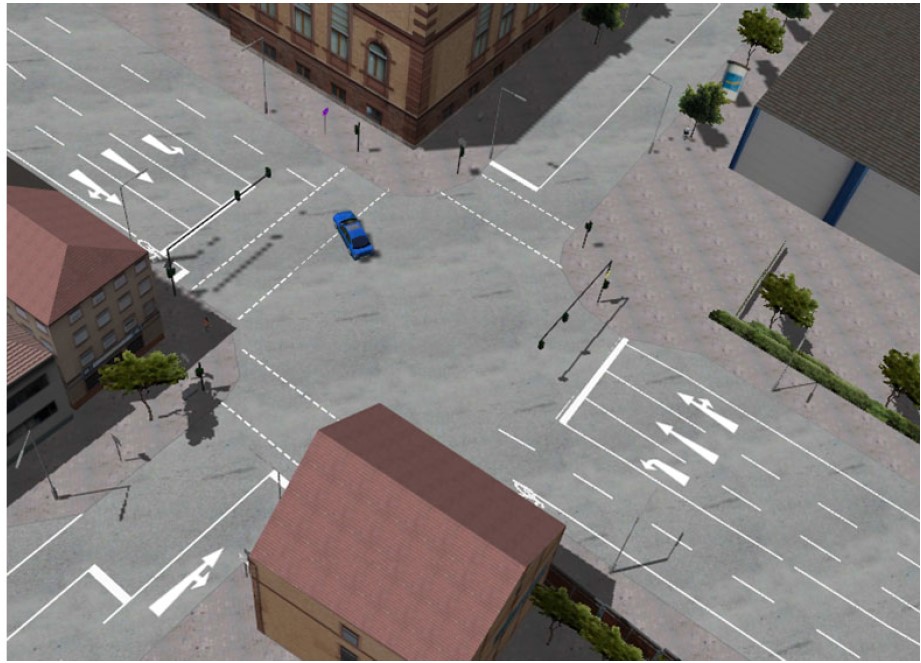}
	\caption{Intersection of the Ko-FAS project \cite{Goldhammer2012}. }
	\label{fig:KoperIntersection}
\end{figure}

\begin{figure}[ht!]
	\centering
	\includegraphics[width=0.6\columnwidth]{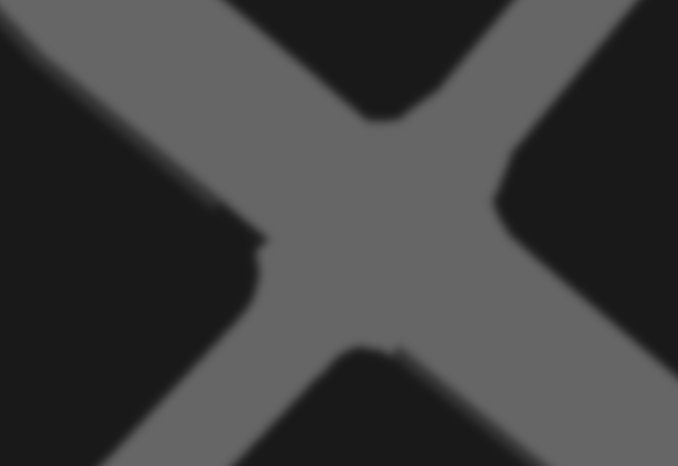}
	\caption{Information Map of the a priori probability of a Bayes classifier for the object class car }
	\label{fig:aPrioriClassProb}
\end{figure}

Among others, contextual information can be a dynamic grid map where static objects are detected. In \cite{Nuss2012} 
an approach to incorporate this information in the preprocessing step of the sensors is proposed, but the same 
approach could also be used to improve the tracking directly, using the Information Map as interface to detection or 
birth probability.
In case of static setups, e.g. intelligent infrastructure, most of the contextual information like traffic lanes, 
sidewalks and much more are static as well. In this case, it is a very practical approach to store context 
information in a static map. One example for such an intelligent infrastructure provides the Ko-PER project, which is 
part of research initiative Ko-FAS \cite{ko-fas2012}, where a major intersection 
was equipped with a network of laser range finders and mono cameras \cite{Goldhammer2012} (see Fig. \ref{fig:KoperIntersection}). 
If a bird eye image or a digital map of this intersection is available, an Information Map can be used to incorporate 
context information. At this intersection, one problem of the tracking using laser range finders is to initialize new 
tracks with a correct orientation angle, because the obtained box-measurements using a box-fit have very uncertain 
information about the orientation. Using an image editing tool, the image of the intersection can be painted in 
different colors, where one color matches one initial orientation. Using this information about the initial 
orientation when instantiating a new track improves the initialization time and precision of the tracks. Another 
example at this intersection is the classification of objects using a Bayes classifier, where the a priori class 
probability can be stored in an Information Map equivalent to Fig. \ref{fig:aPrioriClassProb}. Here the position 
dependent a priori class probability Information Map for vehicles at the Ko-FAS intersection (Fig. \ref{fig:KoperIntersection})
is depicted. The map shows brighter colors where the class probabilities are higher. That corresponds to high class 
probabilities in the areas of the streets where most of the road users are vehicles. The probability declines at the 
curbside and is low at the sidewalks. For each distinguished road user class a map like in Fig. \ref{fig:aPrioriClassProb} 
is needed and the map values of one position have to sum to one.
Even the maximum search radius for a grid based DBSCAN algorithm \cite{Reuter2012b} can be stored as a Map. Further, 
the same approach can be used in dynamic scenarios using the Information Map as an interface to digital maps or a 
databases.

\section{Conclusion} In this contribution Information Maps are presented as a practical approach to determine and 
store position dependent parameters. They are an alternative tool to combine information from different sources 
without using complex analytical descriptions. As shown, the Information Map can be used to provide information about 
parameters in space, context, as well as a priori knowledge. With our approach it is possible to evaluate certain 
parameters, like the detection probability of a sensor, in experiments. These experiments can easily be extended by expert knowledge 
and therefore can lead to better performances. Using Information Maps instead of an analytical description does not 
improve the results necessarily. The convenient parameter representation and the efficient parameter access are the main
advantages of the proposed Information Maps. Especially in case where no analytical description of the parameters
is feasible, like the a priori class probability at intersections, the benefit of 2D representation of parameters using
the Information Maps becomes apparent.

\balance
\bibliographystyle{plain}
\bibliography{References}

\end{document}